# Collision and Diffusion in Microwave Breakdown of Nitrogen Gas in and around Microgaps

J.D. Campbell[1], A. Bowman III[2], G.T. Lenters[1], S.K. Remillard[2,a]
[1]Grand Valley State University, Allendale, MI 49401, USA
[2]Hope College, Holland, MI 49423, USA

## Abstract

The microwave induced breakdown of $N_2$ gas in microgaps was modeled using the collision frequency between electrons and neutral molecules and the effective electric field concept. Low pressure breakdown at the threshold electric field occurs outside the gap, but at high pressures it is found to occur inside the microgap with a large threshold breakdown electric field corresponding to a very large electron oscillation amplitude. Three distinct pressure regimes are apparent in the microgap breakdown: a low pressure multipactor branch, a mid-pressure Paschen branch, both of which occur in the space outside the microgap, and a high pressure diffusion-drift branch, which occurs inside the microgap. The Paschen and diffusion-drift branches are divided by a sharp transition and each separately fits the collision frequency model. There is evidence that considerable electron loss to the microgap faces accompanies the diffusion-drift branch in microgaps.

---
[a] Author to whom correspondence should be addressed. Electronic Mail: remillard@hope.edu

# I Introduction

Being employed commercially and the subject of numerous patents, the application of microplasma is arguably more advanced than the science[1]. Most attention has been on static and low frequency breakdown, with microwave induced microplasma being more recently investigated. The use of microwaves to generate plasma is motivated in part by the near absence of electrode sputtering, which then lengthens the life of the microscopically sized plasma source[2]. Microwave plasma ignition has some key distinctions from static discharge. Electron inertia causes the electron gas to behave as if it were in a smaller *effective* electric field, $E_{eff}$. Also, ion induced secondary electron emission from surfaces is negligible due to the small amplitude of ion oscillation. In both DC and microwave fields, microgap breakdown is known to deviate from Paschen's law[3,4]. The measured threshold breakdown of $N_2$ gas in microgaps, and its distinction from larger gaps, will be described here.

For high frequency breakdown between electrodes in a discharge tube, the similarity law of Lisovskiy, *et al.*[5] conserves the product of pressure and gap size $P \cdot d$ at a fixed value of the product of frequency and gap size $f \cdot d$. They found that at low pressure in large gaps the breakdown voltage is double-valued in pressure because of the possibility of electron loss to the electrodes. Badareu and Popescu[6] also found a double valued electron energy distribution at low pressure. Double valued behavior is not seen however in the case of a small gap discharge inside a large metallic enclosure[7,8], a configuration which more resembles the experiment reported in this paper.

Lisovskiy and Yegorenkov[9] identified several pressure regimes, three of which are evident in our breakdown measurements, each one requiring a unique physical description. The diffusion-drift branch is encountered at higher pressure, so called because diffusion dominates the electron loss and electrons build sufficient kinetic energy through drift to multiply upon collision with neutrals. Breakdown at lower pressure was described in Reference [9] as the "Paschen branch" because, similar to DC breakdown, the kinetic energy available to the electrons upon collision with a neutral molecule increases with decreasing pressure. Finally at very low pressure, multipactor breakdown is evident by a breakdown threshold electric field that is nearly pressure independent since secondary electron emission from the metal surfaces is the only source of new electrons.

Measurement of the threshold breakdown electric field of sub-atmospheric gases in small gaps has some precedent in the literature. Torres and Dhariwal[10] measured a threshold breakdown DC electric field of $2.5 \times 10^7$ V/m in a 24 μm gap for air at 0.2 torr. By particle-in-cell simulation, Radmilović-Radjenović, *et al.*[11] found the threshold for argon inside a 600 μm gap at 100 torr and 2.45 GHz to be $1.5 \times 10^5$ V/m. For a 1 μm gap, they found the threshold to be $1.8 \times 10^8$ V/m. From Iza and Hopwood[12] (computed from Figure 8 in Ref. 12) one can conclude a minimum threshold breakdown electric field for air around 3 Torr and near 900 MHz of about $3.3 \times 10^6$ V/m inside a 45 μm gap, and about $8.3 \times 10^5$ V/m inside a 120 μm gap.

Low pressure microwave breakdown in gaps was previously measured by the current authors using a re-entrant resonant cavity similar to the cavity used here, benchmarked against historical data, and modeled as a collisional process[13]. These experiments have since ventured into microgaps as small as 13 μm[14]. In large gaps, the threshold breakdown electric field $E_{bd}$ in the vicinity of 1 GHz has a minimum around 1 Torr. We will show in this paper that the pressure for a breakdown minimum inside a microgap is much higher.

## II The Experiment

Plasma was ignited in the adjustable gap of a coaxial re-entrant resonator (Figure 1a) excited in a quasi-TEM mode. Using Mylar™ film as a temporary spacer, the gap size, *d*, was set as small as 13 μm. The cone-shaped resonator had a 4 mm diameter flat end which formed a gap with the 4 mm diameter micrometer driven copper tuner rod. Nearly all of the electric field energy in the resonator resides in the extremely uniform electric field between the gap faces (Figure 1b). Resonance was between 0.75 and 1.8 GHz, depending on the gap size, and the unloaded Q was between 1,800 and 2,500. Frequency swept power was generated by an Agilent 8753E vector network analyzer, was amplified to as much as 2 Watts, and was coupled in and out through dipole antennas. At each pressure the microwave power was slowly ramped up until breakdown occurred. Breakdown was confirmed both by observing a sudden drop in transmission through the resonator, and visually through a window in the cavity lid. The coupling coefficients of the resonator were used to compute the dissipated power in the resonator at the point of breakdown. The peak amplitude of the uniform electric field in the gap was then correlated to the dissipated power using a calibration coefficient determined by finite element analysis, using *HFSS*[15] with a pre-seeded meshing tool in the gap. This is unlike Reference [13] which used a perturbation measurement method instead of finite element analysis. This method of finding fields is used routinely in other unrelated works by the authors[16]. Static gas pressure was measured using commercial capacitance manometers.

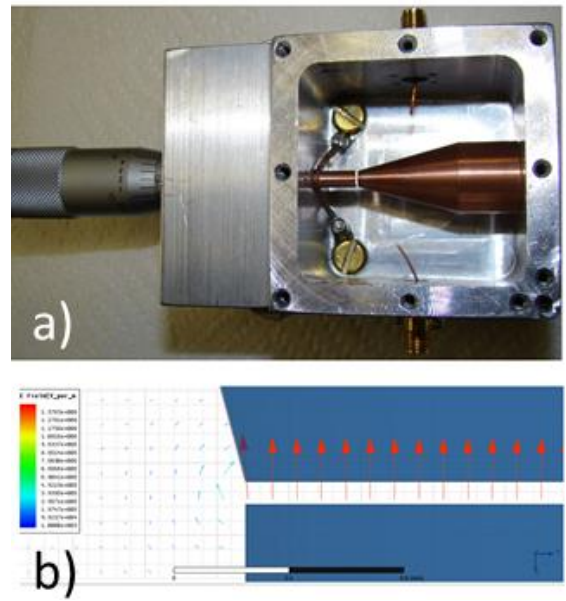

**Figure 1.** (a) The re-entrant resonator includes a copper cone and a 4 mm diameter copper tuner attached to a micrometer. The lid which includes a small viewing aperture above the gap is not shown. (b) The electric field close-up at the 80 μm gap edge computed using *HFSS* shows a high uniform electric field in the gap.

## III Results and Discussion

The effective threshold field $E_{eff,bd} \propto N^m$ depends on the number density $N$ of neutrals and a power law $m$, which measures the extent to which the ion buildup is dominated by collisions[14]. Although questions have recently been raised about its validity at high pressure[17], the effective electric field is expressed as $E_{eff} = E_0 \sqrt{v_c^2/(v_c^2 + \omega^2)}$ where $E_0$ is the applied root-mean-square electric field. The threshold applied electric field at breakdown $E_{bd}$ then varies with pressure $P$ as

$$E_{bd} = CP^m \sqrt{1 + \frac{\omega^2}{(BP)^2}} \qquad (1)$$

where the product $v_c = BP$ is the collision frequency for momentum transfer between free electrons and neutrals. The scale is set by $C$. Fits of Equation (1) to nitrogen breakdown in gaps down to 13μm formed out of a cone and plate geometry were reported in Reference [**14**] at pressures below 30 Torr. Equation (1) fits these breakdown curves well with reduced *chi* square, $\chi_r^2$, ranging from 1 to 3. However, because of the high values of $E_{bd}$ for the smaller gaps, the breakdowns in Reference [14], all measured below a pressure of 30 Torr, were hypothesized to occur outside the microgaps, which is confirmed in this paper.

Figure 2 shows $E_{bd}$ from eight different gap sizes with fits to Equation (1). The higher breakdown electric field with smaller gap size is also evident in data published elsewhere[**11,12**]. Through an opening in the housing, breakdown at low pressure is observed to occur outside the gap. At high pressure, breakdown is observed inside the gap. Equation (1) was fit separately to the low pressure portion and to the high pressure portion. The pressure of the upper minimum depends strongly on $d$, merging with the lower minimum for gaps at and above $d=250$ μm. In the microgaps, the two regions are sharply divided at a gap size dependent transition pressure, $P_t$. A broad single minimum occurs in large gaps (250 to 1,000 μm) with breakdown still inside the gap at high pressure and outside the gap at low pressure. Microgap microwave breakdown simulations by Xue and Hopwood[18] showed that at low pressure, the plasma resides outside the microgap where the electron density is highest. A similar case was shown using larger gaps at 13.56 MHz[8].

J.D. Campbell, *et al.*

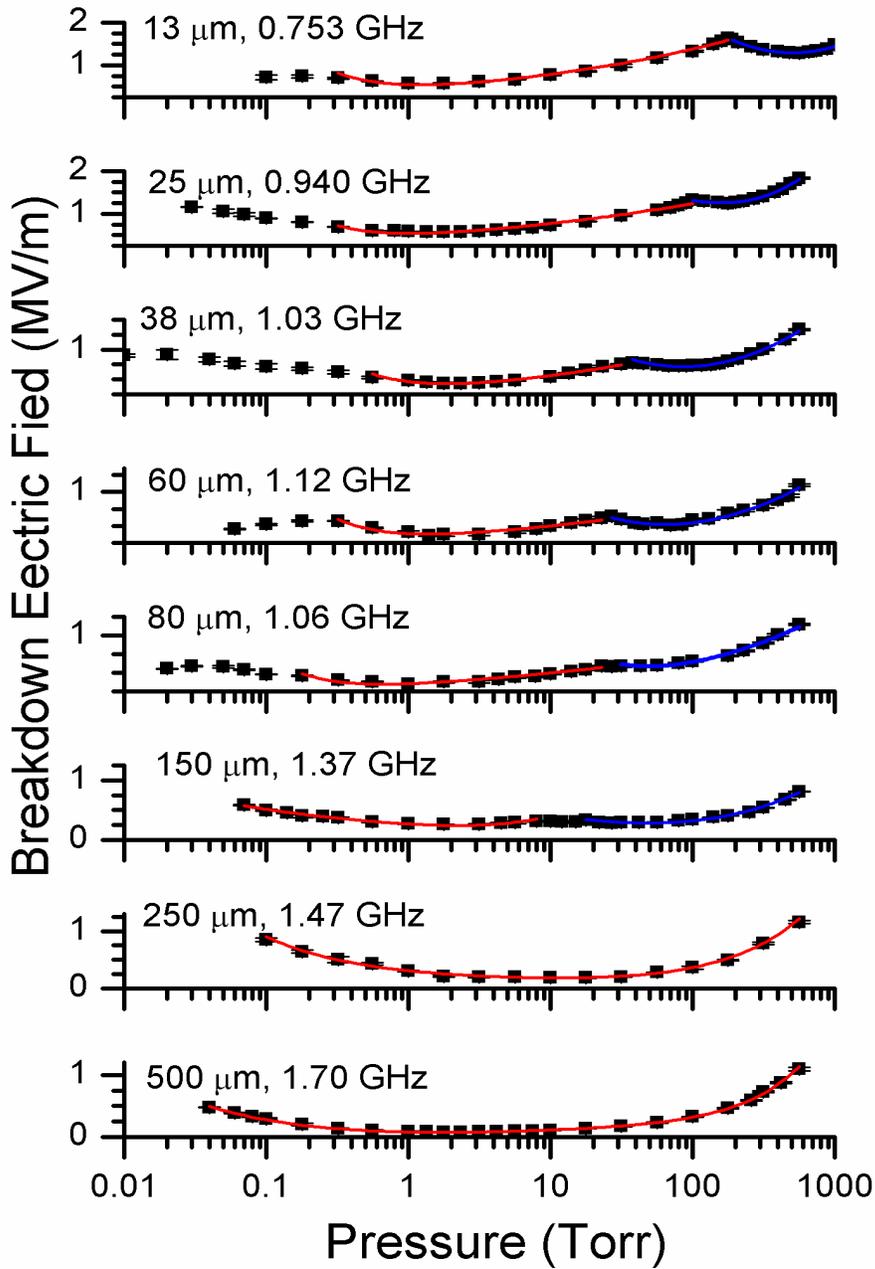

**Figure 2.** RMS threshold breakdown curves for eight different gaps with the gap size and measurement frequency indicated. A sharp transition occurs between breakdown inside and outside the gap, with a clear progression to lower pressures with larger gap. Curves are fits to Equation (1). The two largest gaps are fit to Equation (2).

For the case of a *d*=38 μm gap tested at 1.035 GHz in Figure 2, the electron oscillation amplitude just below the transition pressure of (coincidentally) 38 Torr is approximately[19] $eE_o/m\omega\sqrt{\omega^2+v_c^2}$ =89 μm, relying on the fit to the Paschen branch, with $\chi_r^2$ of 2.0, which gave a collision frequency of $v_c$=6.1[GHz/Torr]·*P* and $E_o$= 7.8x10$^5$ V/m. Throughout the Paschen branch the oscillation amplitude is larger than the gap, but decreasing as pressure rises, until the point at pressure $P_T$ where the oscillation amplitude is similar to the gap, and has been found elsewhere in larger gaps to be approximately half the gap size[9]. The diffusion-drift branch, for the 38 μm gap case, has its minimum at a pressure which is nearly two orders of magnitude higher than the minimum for the Paschen branch. Thus the diffusion-drift branch exhibits a lower collision frequency/Torr, found from the fit to Equation 1, with $\chi_r^2$ of 1.4, to be $v_c$=0.074 [GHz/Torr]·*P*, revealing a large electron oscillation amplitude at $P_T$ of nearly 3 mm. Therefore, in microgaps the transition pressure $P_T$ is that pressure where the Paschen branch oscillation amplitude has reduced to a value similar to the size of the gap and is no longer too large for the breakdown to occur inside the gap. Electron motion inside microgaps above $P_T$ at the threshold electric field is a subject of on-going investigation to be presented in a future paper.

All of the free electron oscillation amplitudes that correspond to fitting the diffusion-drift branch in Figure 2 to Equation (1) far exceed the microgap size. It was shown in the 2004 Northeastern University PhD dissertation of F. Iza[20] that the electron oscillation amplitude at breakdown exceeds the microgap dimension leading to additional electron loss to the gap metal. This added loss in turn leads to an increase in the threshold breakdown voltage of the gap. This offers an explanation for the difference in shapes of the double-minima breakdown curves in Figure 2 from those reported in References [5] and [9] where the diffusion-drift branch is seen at lower values of threshold breakdown electric field than is the Paschen branch. In the large gaps used in those two papers, diffusion is the dominant loss mechanism, positioning $P_t$ at the pressure where the electron oscillation amplitude at $E_{bd}$ equals approximately half the gap size. In microgaps however, electrons are also lost to the metal, adding a large offset to the threshold breakdown electric field, applicable to the diffusion-drift branch only. Whether this offset is additive, multiplicative, or something else, is still a matter of investigation.

By fixing the product *f·d*, Lisovskiy *et al.*[5] showed that the transition pressure $P_t$ between the diffusion-drift branch and the Paschen branch occurs at the same product of *P·d* for all gap sizes. Such a similitude study is not afforded with the current data as no two curves occur at the same value of *f·d*, although confirming this scaling law for microgaps will certainly be a crucial next step in the understanding of microwave microgap breakdown.

With the two breakdown regimes merged in larger gaps, a better mathematical description for 250 μm and above comes from a two-fluid treatment of the pre-breakdown $N_2$ gas. This shunting of the two breakdown branches in Equation (2) is motivated by the observation that with larger gaps which do not exhibit two distinct branches, immediately upon

J.D. Campbell, *et al.*reaching breakdown plasma is observed both inside and outside the gap over a small mid-range of pressures. Figure 3 shows a fit to the hypothesis that $E_{bd}$ is a quadrature summation of threshold breakdown inside and outside the gap

$$E_{bd} = \sqrt{E_{bd1}^2 + E_{bd2}^2} \ . \tag{2}$$

$E_{bd1}$ and $E_{bd2}$ are each individually described by Equation (1), with separate power laws, $m_1$ and $m_2$, collision frequency coefficients $B_1$ and $B_2$, and relative strengths $C_1$ and $C_2$. It was not obvious that the two breakdown conditions should be combined in quadrature, except to note that a poor fit was realized from the function $E_{bd1}+E_{bd2}$ whereas the quadrature function resulted in an excellent fit. Plasma may only be observed over a small. Mid-pressure range (about 1 to 5 torr), however the fit parameter will be such that $E_{bd1}$ is insignificant at high pressures, $E_{bd2}$ is insignificant at low pressures, and Equation (2) then provides a very good description of the large (250 µm<$d$<1,000 µm) gap threshold breakdown electric fields, albeit with more of a mathematical than a physical appeal.

A well-established empirical expression for the threshold breakdown based on diffusion is[21,22]

$$E_{bd} = CP\left[1+\left(\frac{\omega}{\nu_c}\right)^2\right]^{1/2}\left(\frac{D}{P\Lambda^2}+64{,}000\right)^{3/16} \tag{3}$$

where $D \propto 1/P$ is the diffusion coefficient in cm$^2$/s. Although this inverse pressure dependence of $D$ is generally not complete when accounting for diffusion anisotropy[23], it has been found elsewhere to not significantly affect the breakdown condition for parallel plate geometry at 40.68 MHz[24]. We will see that Equation (3) provides a good description of breakdown in the pressure region above $P_t$ with the assumption that $D \propto 1/P$. $\Lambda$ is the effective diffusion length in cm, which depends on pressure[25] as $\Lambda^2 = \Lambda_0^2/\sqrt{1+(P/P_0)^2}$, $P_0$ being a scaling pressure in Torr. Continuing the example of the 38 µm case, a fit of Equation (3) to the diffusion-drift branch yields, with $\chi_r^2$ of 6, a value of $\nu_c$=0.050 [GHz/Torr]·$P$, less than the value of 0.074 [GHz/Torr]·$P$ from the fit of Equation (1), but providing additional confirmation that the high threshold electric field and extremely high pressure of this branch appears to correspond to a low collision frequency per Torr at breakdown. Two distinctions between Equations (1) and (3) are the power law $m$ and the ratio $D/\Lambda^2$. The power law in Equation (1) is unity in Equation (3), as it also is in gap-less microwave breakdown in the open atmosphere[26] and in waveguides[27]. It is perhaps the short effective diffusion length, $\Lambda$, that is found in gaps which causes $m$ in Equation (1) to deviate from unity.

In each microgap, the parameter $P_o$ in the diffusion model of Equation (3) is found to be greater than 10$^{10}$ Torr, indicating that $\Lambda$ is pressure independent and equal to $\Lambda_0$, which for this cylindrical geometry[28] is equal to $d/\pi$. The diffusion coefficient is[26] $D \approx 10^6/P$ (cm$^2$/s), and the term $D/P\Lambda^2$ is thus on the order of $10^{11}$ at 1 Torr for a 100 μm gap, which is within the range of values found when Equation (3) is fit to the data. The quantity 64,000 in Equation (3) is thus insignificant for microgaps, reducing Equation (3) to

$$E_{bd} \approx CP^{5/8}\left[1+\left(\frac{\omega}{\nu_c}\right)^2\right]^{1/2} \tag{4}$$

With the power law $m_2$ for the upper region, shown in Figure 4, scattered around an average value of 0.632, the diffusion model, when reduced to Equation (4), provides a nearly identical description as Equation (1) for breakdown inside the microgap. The diffusion-drift branch breakdown inside the microgap, corresponding to the upper minimum, is treated by Equation (3) as a balance between diffusion loss and ionization. Equation (3), which describes parallel plate geometry with diffusion dominated processes, does not describe the threshold breakdown below the transition, where breakdown occurs outside the microgap. Since Equation (1) fits the low $P$ region with $m$ ranging from as low as 0.15 up to 0.70 there is no chance that Equation (3) could describe the Paschen branch threshold for breakdown outside these microgaps.

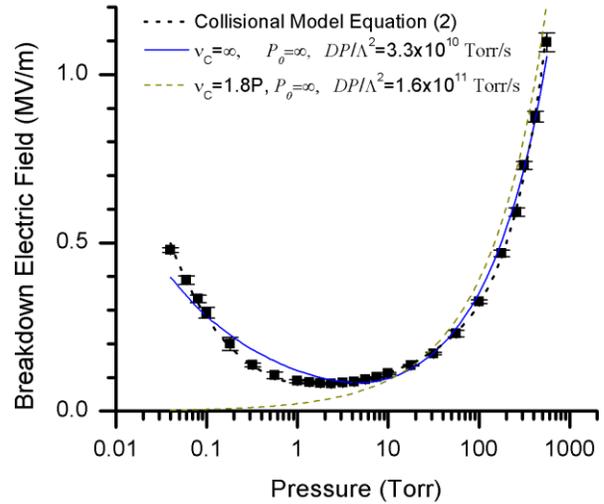

**Figure 3.** Threshold breakdown electric field for a 500 μm gap at 1.7 GHz. The pressure regions for breakdown inside (high pressures) and outside (low pressures) the gap overlap, and the data are better modeled as two simultaneous breakdown processes using Equation (2).

The diffusion model in Equation (3) also fits, albeit with $\chi_r^2$ between 8 and 145, the threshold for the gaps larger than 200 μm, which do not exhibit the double minima. Figure 3 shows two possible outcomes from fitting Equation (3). In one case, the collision frequency, $\nu_c$, is so large that its explicit contribution to the threshold is negligible, and the square root factor in Equation (3) is therefore unity. In the other case, the collision frequency is finite. Both cases fit well at high pressure. So, unless the low pressure data are available, it is not possible to discern the limitation that collisions place on the threshold breakdown. Although less justified physically, the collisional model of Equation (2) which has two additional free parameters fits these "bathtub" shaped large gap curves much better, with $\chi_r^2$ between 3 and 14.

J.D. Campbell, *et al.*

It is evident in Figure (2) that there is a third pressure regime found in gaps smaller than 100 μm and at pressures below about 0.2 Torr. In this higher vacuum regime, the microgap threshold breakdown is lower than the values predicted by equation (1), and a change in mechanism determining the threshold breakdown is clearly evident. This regime, described as the multipactor branch in Reference [9], appears to be influenced by the smoothness and parallelism of the gap faces. The multipactor branch, which is not seen with the larger gaps in Figure 2, manifests neither at a consistent pressure nor to a consistent extent in each gap size breakdown curve. It occurs with a very long mean free path length compared to the gap size and it is dominated by secondary electron emission from metal surfaces.

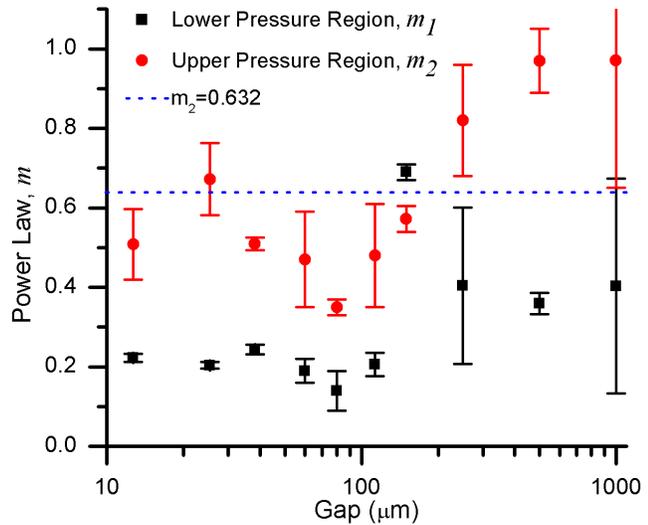

**Figure 4.** Power law *m* for the fits of Equation (1) to the curve below ($m_1$) and above ($m_2$) the transition pressure. Equation (2) was used for the largest three gaps.

Besides modeling $E_{bd}$ with Equations (1) and (4), breakdown in and around large and small gaps can be distinguished by their optical emissions. Spectra taken with an SBIG ST-7E astronomical grade spectrometer reveal that the 1$^{st}$ positive system of $N_2$ is suppressed in a 25 μm gap to within the spectrometer sensitivity, regardless of whether the breakdown occurs inside or outside the gap. In a 500 μm gap, as *P* increases emissions from the 2$^{nd}$ positive system (centered around 400 nm) decrease, while emissions from the 1$^{st}$ positive system (centered around 600 nm) increase, culminating in a yellow nitrogen plasma above 500 Torr.

## IV Conclusion

Microwave breakdown is seen to only occur inside microgaps above a transition pressure $P_t$. Inside these microgaps, the threshold breakdown model derived solely from collisions using the effective field concept converges on the model that includes diffusion. However, the electron oscillation amplitude prior to breakdown inside the microgap is much larger than the gap size, adding an offset to the threshold breakdown electric field. With Paschen branch breakdown, at least when *d*≤150 μm, it was found that $E_{eff,bd} \propto N^{0.2}$ and Equation [3] does not describe the threshold. It is unclear whether this results from gap geometry or collisional processes dominating over the diffusive processes, or perhaps, a mixture of the two.

This work was supported by NSF Grant PHY/DMR/1004881 and by a Jacob E Nyenhuis Faculty Development Grant from Hope College. Valuable input came from Prof. Peter Gonthier of Hope College. Significant insight was gained from the reviewer's comments.